# Complex forming behaviour of α, β and γ-cyclodextrins with varying size probe particles *in silico*

N.R.M. Nelumdeniya and R.J.K.U. Ranatunga*

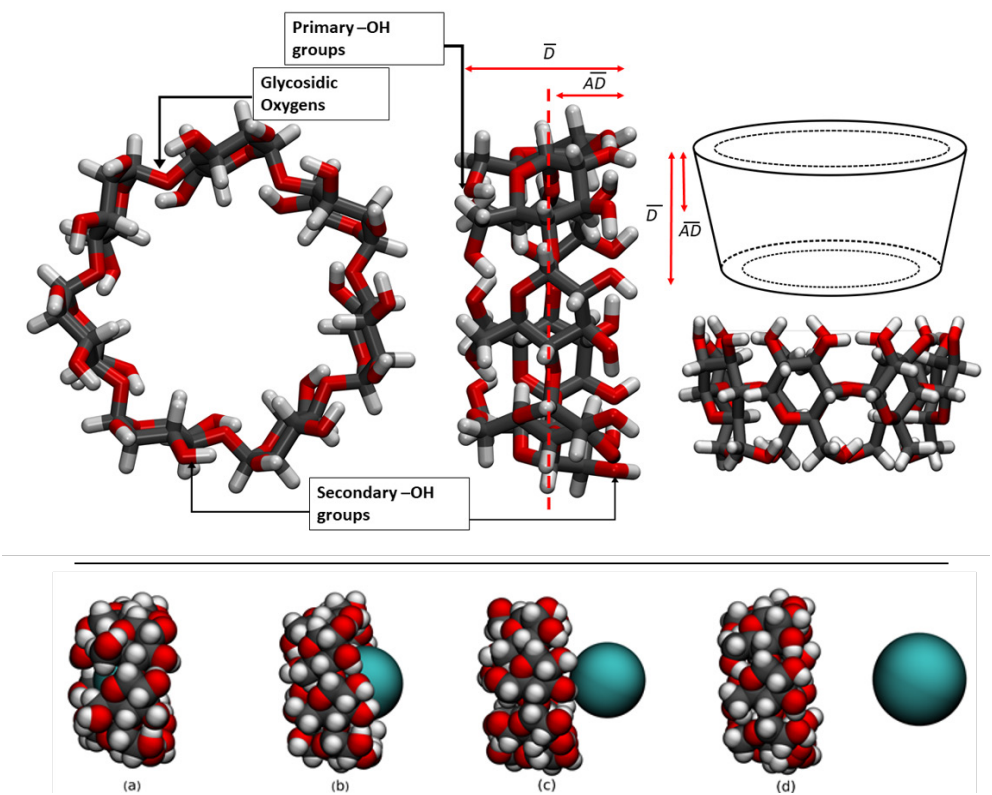

# Highlights

- Cyclodextrins (CDs) can hold considerably larger guests than their known cavity size even though they are considered to have a rigid truncated-cone structure.

- CDs are somewhat flexible; they distort to embrace the guest, and the complexes formed are in dynamic equilibrium.

- CDs can successfully complex intermediately hydrophilic particles confirming that the cavity is not so hydrophobic.

- Larger particles with intermediately hydrophilic nature closely contact the larger ring of the CD-ring making non-inclusion associations.

**RESEARCH ARTICLE**

# Complex forming behaviour of α, β and γ-cyclodextrins with varying size probe particles *in silico*

**N.R.M. Nelumdeniya[1,3] and R.J.K.U. Ranatunga[2,3*]**

[1]*General Sir John Kotelawala Defence University, Ratmalana, Sri Lanka.*
[2]*Department of Chemistry, Faculty of Science, University of Peradeniya, Peradeniya, Sri Lanka.*
[3]*Postgraduate Institute of Science, University of Peradeniya, Peradeniya, Sri Lanka.*

**Abstract**: Cyclodextrins (CDs) are cyclic oligosaccharides composed of glucopyranose units bonded together to form a truncated cone that can make inclusion complexes with guest molecules. The α, β, and γ-CDs, which respectively comprise six, seven or eight glucopyranose units, are used extensively in pharmaceutical formulations as functional excipients. The cavity sizes of all three natural CDs have been approximated using static structures but a growing consensus is that the CDs are flexible; moreover, the size range of molecules that CDs can accommodate has not been systematically studied. Here the results of molecular dynamics simulations performed using spherical continuum probe particles of different sizes to observe the complex-forming behaviour of CDs are presented. Results revealed that CDs can make dynamic complexes with guest molecules that are larger than their reported cavity sizes. Probe particle with intermediate hydrophilicity ($\epsilon = 0.2$ kcal mol$^{-1}$), with nominal radius in the range of 0.5 - 1.1 Å (effective radius of 2.61 - 3.41 Å), makes the complexes with α-CD. For β-CD, these values range from 0.9 - 1.9 Å (3.75 - 4.26 Å) and for γ-CD 1.4 - 2.8 Å (3.72 - 5.09 Å) respectively.

## INTRODUCTION

Excipients have many roles in the pharmaceutical industry. They can modify physical, chemical and biological characteristics of drugs, leading to improved product characteristics. Excipients that form inclusion complexes with drug molecules can improve the solubility, stability, and even mask the taste of the drug and imbue many other advantages. Inclusion complexes are molecular associations involving only non-covalent bonds in which one molecule known as host possesses a cavity that can admit 'guest' (drug) molecules (Brown *et al.*, 1991). Cyclodextrins (CDs) are a class of carbohydrates that are known to form inclusion complexes and hence are of pharmaceutical interest as hosts for developing multifunctional nanostructures (Cova *et al.*, 2019), to enhance the solubility, stability, safety and bioavailability of drug molecules (Rasheed *et al.*, 2008).

### Cyclodextrins (CDs)

Cyclodextrins (CDs) are cyclic oligosaccharides with the arrangement of glucopyranose units with 1-4 linkage forming a truncated cone that can trap or encapsulate other molecules (Loftsson and Brewster, 1996). The three natural CDs (α-, β- and γ-CDs) are built up from 6, 7, and 8 glucopyranose units, respectively (Bibby *et al.*, 2000), and share similar shapes but different physicochemical parameters (Figure 1). The arrangement of glucopyranose units forms a truncated cone consists of a wider ring and a narrow ring (Cheng *et al.*, 2011), and the inner lining of the cavity is covered with hydrogen atoms and glycosidic oxygens. The non-bonded electrons on these oxygen atoms impart some Lewis basicity to the cavity and give rise to the uniqueness of the CD molecule, where it has a hydrophilic outside, which is soluble in water, and a less polar cavity described as a 'micro heterogeneous environment', with hydrophobicity similar to an aqueous ethanolic solution (Loftsson and Brewster, 2010; Rodrigues *et al.*, 2011; Jambhekar and Breen, 2016). However, it should be emphasized that this hydrophobicity of the cavity is always relative to the physicochemical nature of the guest molecule and the surrounding environment. Appropriately sized molecules and macromolecules can be housed via strong, yet reversible associations (Tan *et al.*, 2014). The primary hydroxyl groups of the sugar residues at the narrow edge and the secondary hydroxyl groups at the wider edge are orientated to the cone exterior (Uekama *et al.*, 1985). Secondary hydroxyls form strong hydrogen bonds rigidifying the wider edge of the CDs while primary hydroxyls are free to rotate. This freedom of rotation can reduce the diameter of the narrow edge and giving CD a truncated cone-like shape (Przybyla *et al.*, 2020).

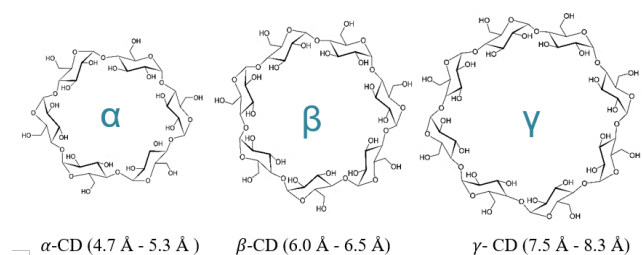

α-CD (4.7 Å - 5.3 Å)    β-CD (6.0 Å - 6.5 Å)    γ- CD (7.5 Å - 8.3 Å)

**Figure 1:** Structures and approximate geometric dimensions of α-CD, β-CD and γ-CD.

It is known that β-CD has a rather rigid structure; but the lowest water solubility of all CDs is due to the H-bond belt formed among the hydroxyl groups of the adjacent glucopyranose units. The H-bond belt is incomplete in the α-CD molecule, due to a distortion of one glucopyranose unit. On the other hand, γ-CD has a more flexible non-coplanar structure, and it is the most soluble of the three natural CDs.

**Complex formation process**

The parameters, such as geometric compatibility, structure, charge and polarity, are important characteristics of the guest which participate in the complex formation though few studies have examined these factors systematically. Since the functional groups, polarity, structure and shape vary with the molecules, no general findings regarding the size range of the guest molecules that CDs can accommodate is available in the literature. The results obtained from this study give an opportunity to re-evaluate the cavity sizes of CDs devoid of influence of above factors (of guests).

In aqueous solution, polar water molecules fit into the lipophilic cavity of CDs; but these can be replaced by a more favoured guest molecule which is less polar (Szejtli, 2004). It is believed that the CDs are capable of forming inclusion complexes with many drugs by taking up a whole drug molecule, or rather some non-polar part of it, into the cavity (Másson *et al.*, 1999). Multiple non-covalent forces such as London dispersion forces, hydrogen bonding, dipole-dipole interactions and hydrophobic interactions act synergistically to form a stable complex. Some suggest that even the release of high-energy water from the CD cavity, relief of structural ring strains in the CD molecule, and changes in solvent surface tension as well as polar and steric factors can further contribute to complexation. However, the relative contributions and even the specific nature of the different forces involved in the process are still not well known (Faucci *et al.*, 2002). Moreover, a force responsible for complexation for one series of molecules may not be responsible for another series of molecules, causing difficulty in predicting how well a particular molecule might bind with a CD. Even having only the lipophilic parts of the molecules are questionable.

No covalent bonds are involved in drug/CD complex formation, and in aqueous solutions, the complexes are readily dissociated, and free drug molecules are in rapid equilibrium with drug molecules bound within the CD cavity (Másson *et al.*, 1999).

There are several experimental and theoretical studies on complexation behaviour of CDs. However, a detailed description of the effect of CD cavity size and conformation on the formation of inclusion complexes, especially at the atomic level, is still lacking (Cova *et al.*, 2019). Further, even though this host-guest chemistry has been known for some time, CD-based formulations have mainly been developed through a trial-and-error approach which is time and resource-consuming, and this empirical approach does not help to understand the process of CD and the structure-activity relationships between the CD and guests.

Molecular Dynamics (MD) simulations are one of the most versatile and widely applied computational techniques for the study of biological macromolecules and drug-like compounds and can be used to understand the microscopic origin of physical properties of a chemical process or to predict the qualitative changes of a reaction by changing different parameters that cannot be tested experimentally (González, 2011). Hence, they can be used to accelerate the drug development process while minimizing the time and financial burden associated with conventional experiments.

This MD study aims to investigate the 1:1 complexation of guests with CDs. Investigation of the size of guest molecules that a CD can accommodate was motivated by the fact that current measurements of cavity sizes of CDs are only through static molecular structures. The flexibility of CD molecules depends on the intramolecular hydrogen bonds that they can make (Charumanee *et al.*, 2006; Parmar *et al.*, 2018), and the guest molecules can change the cavity size distorting the glucopyranose rings depending on the functional groups of the guest molecule and the nature of the surrounding environment (solvent, temperature, etc.). Therefore, this study was aimed to measure the size range of the guest molecules that CDs can accommodate by varying the size of a generic probe particle devoid of any functional groups and to generalize the findings to guide prediction, and future drug development by selecting the best-fit size of the intermediately hydrophobic guest molecule to the CDs cavity.

**MATERIALS AND METHODS**

**Continuum model**

The computational expense of using MD to simulate bulk or nanostructured materials can be reduced by applying continuum models, which can capture properties while maintaining computational efficiency (Huang and Sun, 2007; Gauthier *et al.*, 2019). In the continuum approximation, the individual sites of a solid are smeared out throughout the material. The total interaction between the solid particle and atom can then be approximated by integrating the total interactions. This concept has been documented and developed in several research papers, and has been implemented in several ways (Hamaker, 1937; Henderson and Plischke, 1986; Girifalco, 1992; Ranatunga and Nielsen, 2011).

Here a spherical continuum particle was used to probe the size of guest molecules of CDs. This model can remove the influence of varying geometries of molecules (if used) on the complexation, which is beneficial in determining the best-fit size. Further, the size can be changed continuously rather than incorporating multiple functional groups with different atoms, and finally, the exact shape allows calculating the geometric positioning within the cavity accurately. Furthermore, since no different functional groups are present, the hydrophilicity can be set to a fixed value for different sizes.

For two atoms, $i$ and $j$, at distance $d_{ij}$ the LJ interaction potential can be given by,

$$U(d) = 4\epsilon_{ij}\left\{\left(\frac{\sigma_{ij}}{d_{ij}}\right)^{12} - \left(\frac{\sigma_{ij}}{d_{ij}}\right)^{6}\right\} \text{ (Zhen and Davies, 1983)}$$

Here $\epsilon$ is the strength of the interaction or the depth of the well, and $\sigma$ is the exclusion volume. Both parameters depend on the type of atoms and hence they are indexed as such. As discussed above, the total number of interaction sites can be replaced with a site density $\rho$ and the total interaction between the solid particle and atom $i$ can now be approximated with an integral. For a spherical solid particle with radius $r$, this integral take the form of

$$U(R;r) = \int_0^\pi d\theta \int_0^{2\pi} d\emptyset \int_0^r a^2 \sin\emptyset \, da \rho U(d)$$

where the $d$, the distance between the atom and a point in the sphere can be defined through geometry [Figure 2; (Chiu *et al*., 2009)]. The analytical solution to the integral and the shape of the potential is given below:

$$U(R;r) = \frac{16\pi\rho\epsilon\sigma^{12}r^3(5r^6 + 45r^4R^2 + 63r^2R^4 + 15R^6)}{45((R-r)^9(R+r)^9)} - \frac{16\pi\rho\epsilon\sigma^6 r^3}{3(R-r)^3(R+r)^3}$$

The integral was validated using a diamond nanosphere with the same size and site density of the continuum particle and the results are shown in Figure 3. As shown, both follow a similar trend in potential.

**Defining the hydrophilicity**

To find a suitable $\epsilon$ for the hydrophilicity of the probe, a series of simulations were carried out embedding the appropriately sized probe particles with different values in a cell containing heptane and water. Then the density distributions (Figure 4) were analyzed, and the value of 0.2 kcal mol$^{-1}$ considered as an intermediately hydrophilic one where the particle lies on the interface.

**Selecting the particle sizes for the simulation.**

Since CDs are in high tendency to make complexes with guests in the size range of fit or nearly fit range, literature was used as a rough guide, and probe particles were generated expanding the range of sizes investigated. It is reported that $\alpha$-CD has a cavity size of 4.7 Å - 5.3 Å, $\beta$-CD of 6.0 Å - 6.5 Å, while $\gamma$-CD is 7.5 Å - 8.3 Å (Loftsson and Brewster, 1996). However, most of these sizes were measured using the static molecular structures created by modeling software even though the CDs are flexible molecules. Therefore, to improve the accuracy and the versatility, simulations were conducted starting from the particle size with a radius of 0.5 Å (1.0 Å diameter) to larger values to find the optimum size of the particle which each CD can accommodate.

**Details of the simulation**

All simulations were carried out using NAMD (Phillips *et al*., 2005) 2.10 molecular dynamics software using TIP3P water model where the system components were represented using the CHARMM 36 topologies and interaction parameters (Vanommeslaeghe and Mackerel, 2015). The van der Waals forces were modelled using Lennard-Jones potentials, which were then tabulated for interactions involving the probe particle and calculated explicitly for all other interactions, with a distance cutoff of 12 Å, switching smoothly from 10 Å. The strength of interactions of $\sigma$ and values of atom type and continuum probe particle pair was calculated using the Lorentz-Berthelot mixing rules (Lorentz, 1881). Pair-lists were calculated within a radius of 13.5 Å and updated every two time steps. Systems were initially minimized for 500 steps using the steepest descent algorithm and then simulated with Langevin dynamics with a pressure of 1.01325 Pa, at a temperature of 300 K and a time step of two femtoseconds (2 fs) using the RATTLE and SHAKE algorithm to constrain hydrogen vibrations.

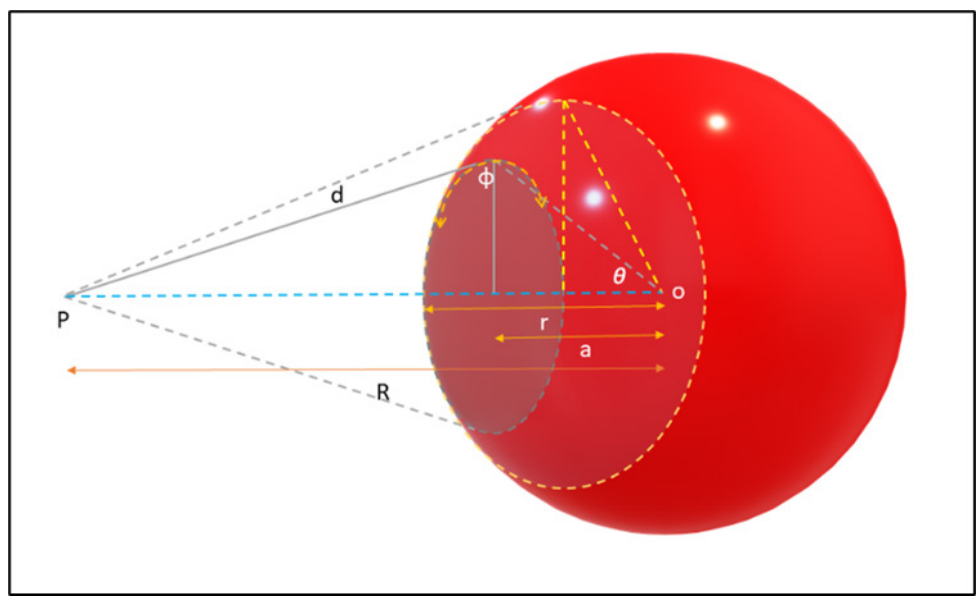

**Figure 2:** The total potential of an atom (at point P) with a solid particle (at origin O) can be found by integrating the site potential at a point (a, θ) within the sphere. Here R is the distance between the outside atom and the center of mass of the probe particle.

For the probe particle interaction, continuum probes with 0.5 Å to larger were generated with radius increments of 0.1 Å and were subjected to 10 ns simulations separately for each α, β and γ CDs. Even though the nominal radii of the continuum solid particles are set, the actual effective sizes of those depend on the interaction potential. To determine the effective radii, a series of simulations were carried out by setting the particles in water and tabulating the radial pair distribution function (RDF).

## Analysis

MD trajectories were analyzed using VMD (Humphrey *et al.*, 1996) through inbuilt and custom coded scripts. To calculate the complex forming behaviour quantitatively, average planes of the CD cavity were defined using the positions of the glycosidic oxygens and primary and secondary hydroxyl oxygens of the CD molecules (Figure 5). Depending on the distance of the probe particle's center of mass (COM), from the average center plane, the level of the inclusion complex was determined.

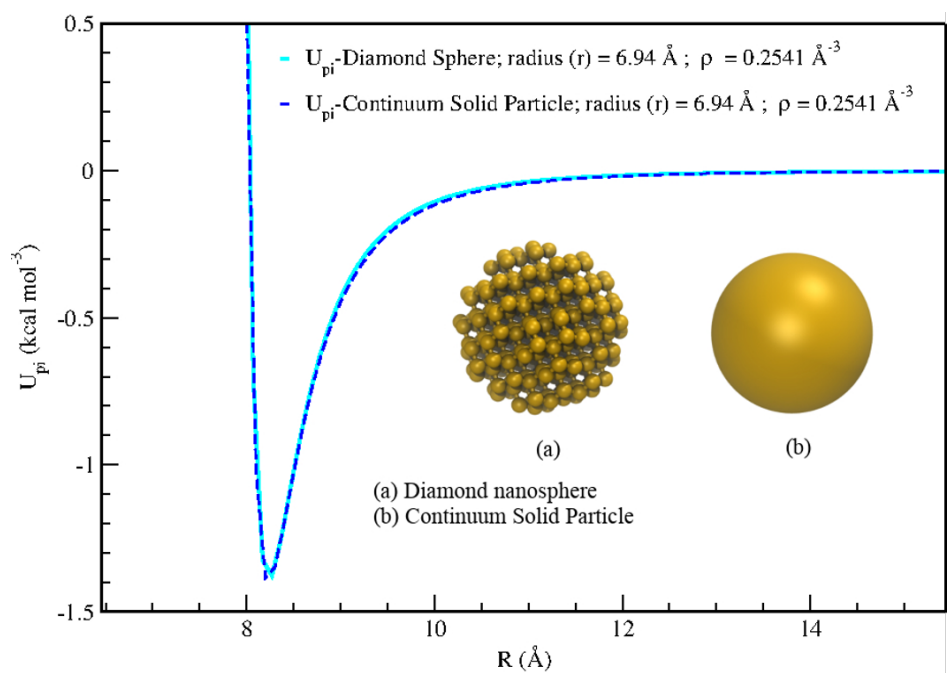

**Figure 3:** Comparison of the potentials obtained from the diamond nanosphere and the continuum solid particle. The site density of nano diamond is calculated from the unit cell volume ($\rho$ = 0.2541 Å$^{-3}$).

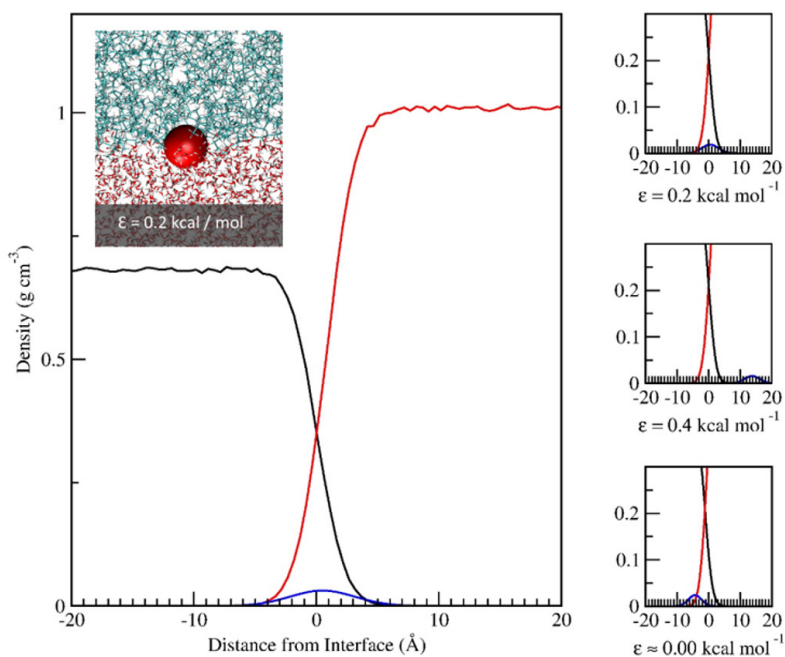

**Figure 4:** Density distribution plots showing the location of the probe particle, for different hydrophilicities ($\epsilon$∼ 0.0 kcal mol$^{-1}$, $\epsilon$ = 0.2 kcal mol$^{-1}$, $\epsilon$ = 0.4 kcal mol$^{-1}$) values.

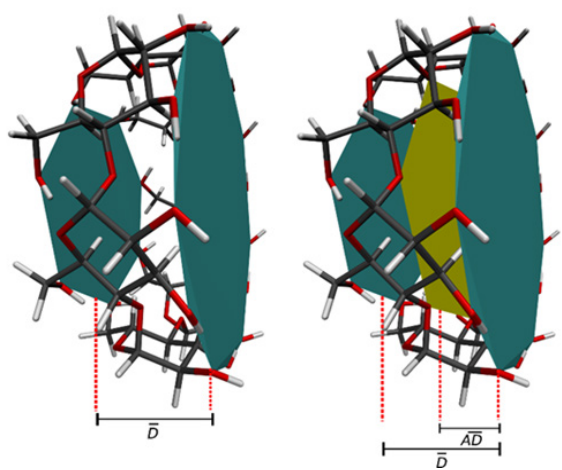

**Figure 5:** Depth of the cavity is defined as D, a region between edge planes of the CD cavity; the complexing region (accessible depth) can be defined as AD, the region between planes created by glycosidic oxygen and edge plane of the larger ring.

### Defining the level of complexation

The perpendicular distance from the central plane to the COM of the guest molecule (probe particle) is defined as **d**. The distance between the average middle plane to the plane at the edge of the wider ring is hereinafter named as the 'accessible depth' (**AD**).

Since the CDs are flexible, the AD changes with time, but the average values from MD trajectories for accessible depths are 2.342 Å for α-CD, 2.239 Å for β-CD and 2.142 Å for γ-CD. The particle may form complexes or can show associated behaviour in distances greater than the above values. The best fit sizes can be identified by looking at the average distance from the middle plane and the radius of the particle. The complexing behaviour was classified as below.

$d + r \le AD$ → type **(a)** complexes - fully included form

$d \le AD$ → type **(b)** complexes - partially included form

$d \le AD + r$ → type **(c)** loosely associated

$d > AD + r$ → type **(d)** dissociated

For the characterization of the binding, each saved point in the trajectory was classified as either **a, b, c** or **d** form. If over 80% of the trajectory showed either complexation (a or b) then the respective size and hydrophilicity were considered favourable for complex formation. As shown in Figure 6, if the complexation of the probe particle is favourable, the CD will make full inclusion complexes [Figure 6(a)]. In some cases, it will make partial complexes [Figure 6(b)] or remain uncomplexed but loosely associated [Figure 6(c)]. If the complexation is not favoured at all, the guest may remain dissociated [Figure 6(d)]. Further, average distance distributions were analyzed by plotting the distance of the particle from the average plane in every frame of the trajectory.

### Effective particle size

The radius of the particle is the distance at which the surface nuclei are from the centre. However, the potential made by the probe particle and the electron cloud of any external molecules creates a void distance where they cannot approach each other and that is the effective size of a particle. To determine the effective size of particles, an inbuilt plugin in VMD to check the Radial pair Distribution Function (RDF) was used. Figure 7 shows the changes in RDF, giving effective interactive sizes of the particles with a nominal radius of 3.10 Å and $\epsilon = 0.2$ kcal mol⁻¹. The effective radius (eR) can be taken as the point of closest approach of the solvent, which is practically found as the distance corresponding to the 0.1 in the first peak of the RDF (Figure 7). Quantitatively that is 5.38 Å for the 3.10 Å radius probe particle.

The linear relationship between the actual radius and the effective radius was found to be

$y = 1.00065x + 2.28616$ and is shown in Figure 8.

### RESULTS AND DISCUSSION

Due to the size, functional groups and other interactions with guest molecules, CDs may encapsulate the whole guest molecule or a part of it. However, a force responsible for complexation for one series of molecules may not be in effect for another, causing difficulty to predict how well a particular molecule might bind with CDs. It is believed that the guest size affects CD-guest contact and the inclusion degree, inducing CD deformation of certain level, which opposes inclusion event (Cova *et al.*, 2019). Here a novel methodology is described to investigate the optimum size of probe particles to form stable complexes with all three natural CDs using a continuum model.

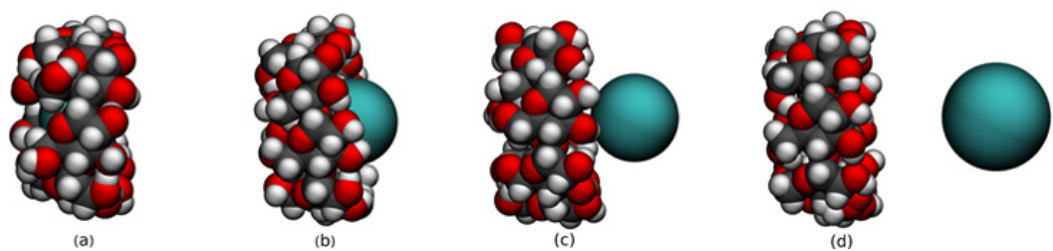

**Figure 6:** (a) Full inclusion complex (b) Partial inclusion complex (c) Loosely associated, (d) dissociated.

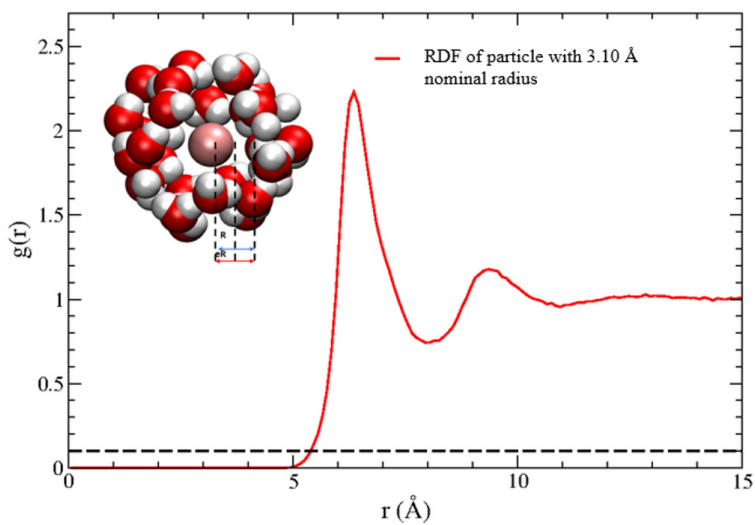

**Figure 7:** Change of the Radial pair Distribution Function (RDF) of particle with 3.1 Å  nominal radius.

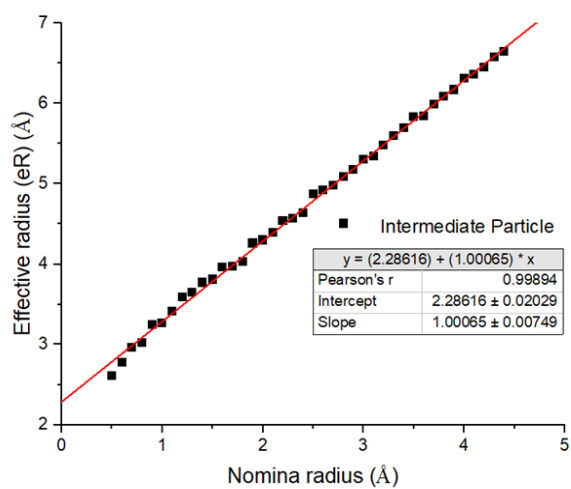

**Figure 8**: The relationship between the nominal radius and the effective radius.

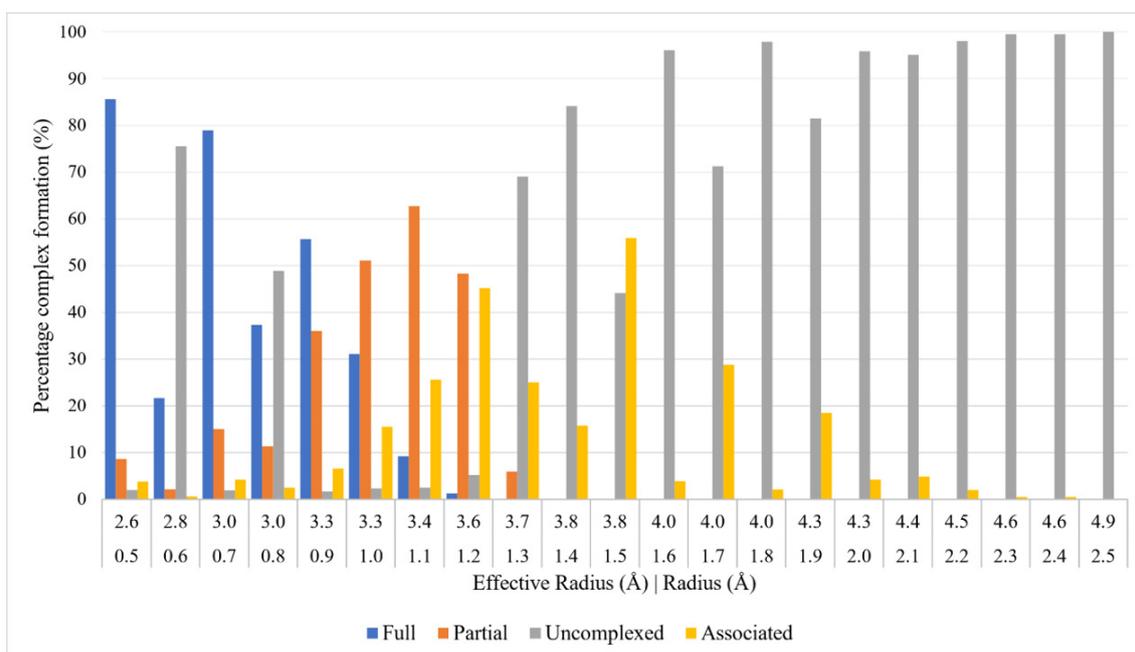

**Figure 9:** Interaction between $\alpha$-CD and probe particle; percentage of complex formation.

As discussed in the introduction and in the methodology, the inclusion complexes formed between CDs and guests are driven by non-covalent interactions and the binding is reversible, yet the complexes are dynamic; constantly forming and dissociating. This behaviour was observed in the trajectories resulting from the MD simulations. For ease of analysis, four (4) major binding states of CD-guest complexes with 1:1 stoichiometry was designated.

The polarity of the cavity has been estimated to be similar to an aqueous ethanolic solution (Brewster and Loftsson, 2007). Hence the probe particle with $\mathcal{E}$ = 0.2 kcal mol$^{-1}$ was chosen due the match in hydrophobicity.

### Interaction between α- CD and continuum particle

The reported cavity diameter of the α-CD is 0.47-0.53 nm (4.7-5.3 Å) (Jambhekar and Breen, 2016). However, as previously stated, due to the flexibility of the molecule and the physicochemical influence of the guest molecule and the surrounding environment, the actual accessible cavity size may be different from the reported value. In our results, the maximum diameter of probe particles which α-CD can accommodate is above the reported (static) cavity size. Figure 9 shows the percentage variation of the binding state of probe particle with its size.

The average AD of the α-CD is 2.342 Å. The radius range of 0.5-1.1 Å (predicted eR 2.61 - 3.41 Å) is the feasible size where a probe particle can make a successful inclusion complex with α-CD. However, in this range, probe particles with radius between 0.5 and 0.7 Å can make full inclusion complexes while others show a mixture of full and partial complexes throughout the trajectory. Radius above 1.3 Å will not show complexed behaviour. The maximum radius where it can make a full inclusion complex is 0.7 Å with the predicted eR of 2.96 Å. Interestingly some sizes show association at a considerable percentage.

To further characterize this complexation behaviour, trajectory points classified as complexed were analyzed, and a histogram of the distance (**d**) between the average plane of glycoside oxygens, and the center of mass of the particles was generated. The histograms were close to normal distributions, indicating that probe particle complexes under a restorative potential, meaning that two forces compete: one pushing the particle inside the CD cavity, and the other pushing it further away. The histograms were fitted to standard normal distributions through non-linear curve fitting. Figure 10 shows the normalized curve obtained from the fitting.

It must be emphasized that the normal distributions were generated only for full, or partial inclusion complexes and associated conformations for all three CDs since in the un-complexed state particles are dispersed at random distances.

It is observed that the distributions move towards larger **d** values, *i.e.*, the larger ring of the CD cavity with increasing particle size. This behaviour can be seen with all three types of CDs that we analyzed. Furthermore, the width of the distribution (variance) indicates how strong the complexing behaviour with continuum particle is. If the distribution is narrow, it is the state where the complexation is strong, and the particle is localized in the cavity with minor fluctuation in its location. But when the distributions are broad, this means the bonding is weak and the particle is more dynamic. This also agrees well with the accessible depth (AD); as noticed in Figure 11, the particles which show distributions with large portions outside of the AD, are also broader and indicate partial complexation.

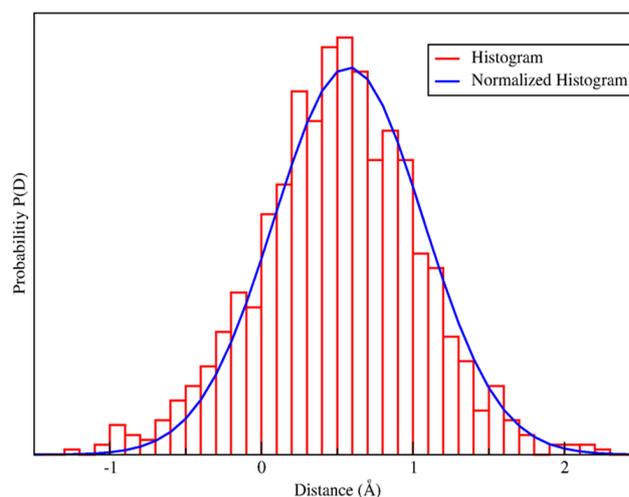

**Figure 10:** Example of a distribution histogram of probe particle - CD distance. The histogram was fitted locally to a normal curve to obtain the mean distance.

With these results, it is noticed that the α-CD can hold much larger probe particles, and this behaviour is expected since the chemical nature of the CD cavity is not purely hydrophobic but rather equal to the aqueous ethanolic solution (Loftsson and Brewster, 2010; Rodrigues *et al.*, 2011; Jambhekar and Breen, 2016). Further, since the CD molecule is flexible it has some ability to expand or shrink with the influences made by the outer environment and the interacting guest molecule.

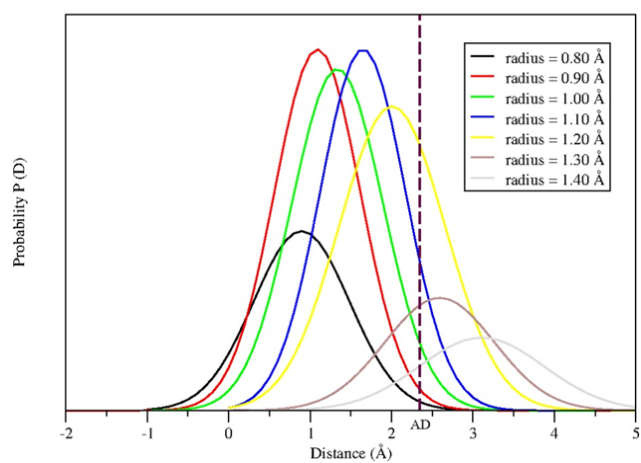

**Figure 11:** Normalized distributions of the distance (d) between the defined plane of α-CD and probe particle.

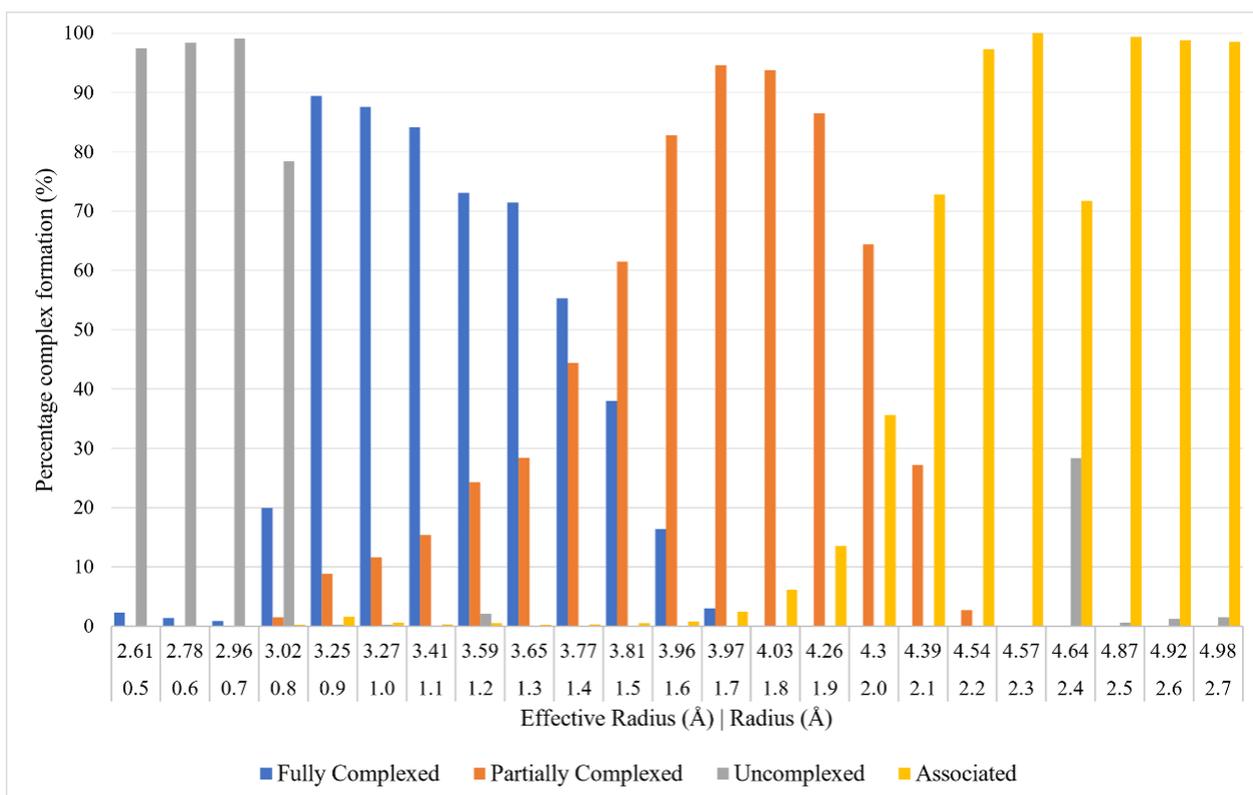

**Figure 12:** Interaction between *β*-CD and probe particle; Percentage of complexation.

**Interaction between *β* - CD and continuum particle**

It is believed that the molecular dimensions of the *β*-CD cavity (diameter 0.60-0.65 nm) is   ideal for inclusion complex formation with drugs (Szente *et al*., 2018). Several drugs have been formulated with *β*-CD to improve their physicochemical parameters.

In our study, the AD of the *β*-CD was found to be 2.24 Å. Sizes of the probe particles obtained are greater than the diameter of the CD cavity as approximated by computer design previously. This is mainly due to the flexibility of the CD molecule when it is going to interact with the guest (Fujita *et al*., 1999; Ye *et al*., 2016), and it is explained that the cyclodextrins offer highly attractive hosts for mimicking the dynamic induced-fit model like enzyme action. Even the distortion of glycosidic units is possible. The distribution of **d** is given in Figure 13. Interestingly, for small sizes, even if they make full inclusion complexes, the particles can move freely due to the higher freedom of movement resulting in broader distributions. However, when increasing the size, this movement gets restrained, and particles tend to be confined to a smaller region.

Probe particles with a radius of 0.9-1.9 Å range (predicted eR 3.25-4.26 Å) make the complexes in type **a** and **b** modes. A higher percentage of fully included form can be seen in 0.9-1.1 Å (3.25-3.72 Å) (Figure 12). Complexes are dynamic in nature. Here the guest particle size range is greater than the *α*-CD since *β*-CD has a larger cavity size due to the additional one glucopyranose unit in the ring. The particles showing weak association is also shifted to larger sizes even though the guest probes cannot make full or partial complexes. This indicates that CDs can make non-inclusion complexes where possible. It is well known that that CDs can arrange like micelles when larger guest molecules are present in the environment, however we limit ourselves to 1:1 stoichiometry in this work.

As shown in Figure 13, the degree of freedom is limited for larger sizes that make full or partial complexes. Therefore, the average fluctuations from the mean distance are limited, causing narrower distance distributions. However, when they make partial complexes, the distributions are broadened again and the freedom in movement is restored.

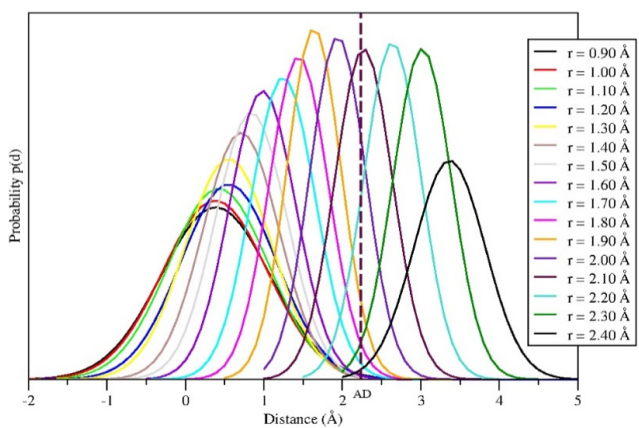

**Figure 13:** Normalized distributions of the distance (d) between the defined plane of *β*-CD and probe particle.

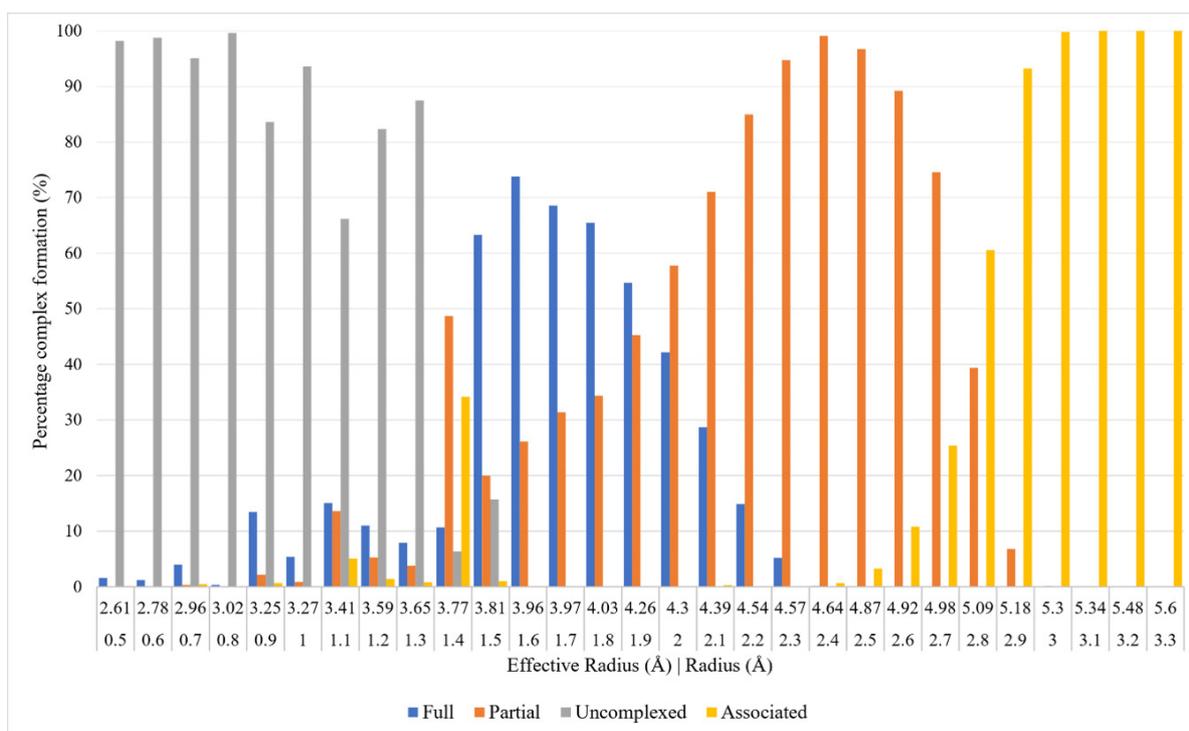

**Figure 14:** Interaction between γ-CD and probe particle; percentage of complexation.

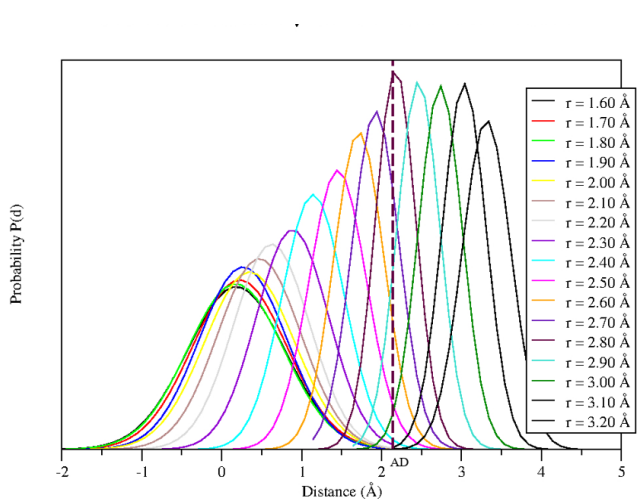

**Figure 15:** Normalized distributions of the distance (d) between the defined plane of γ-CD and probe particle.

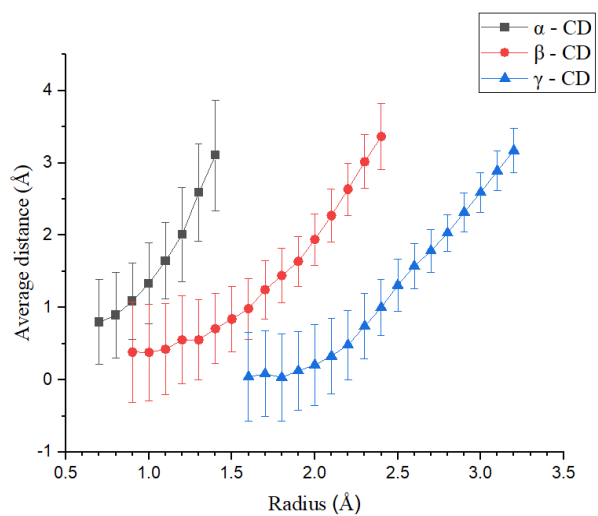

**Figure 16:** Average distance change with the nominal radius of the particle.

### Interaction between γ-CD and continuum particle

Of all three nature derived CDs, γ-CD is the largest. Its cavity diameter is reported as 7.5-8.3 Å in previous literature (Jambhekar and Breen, 2016). Upon analysis, the ensemble averaged AD of the molecule was found to be 2.14 Å.

Radius in the range of 1.4-2.8 Å (predicted eR range 3.72 - 5.09 Å) probe particles can make inclusion complexes with γ-CD but not in the fully complexed state but having rather a mixture of full and partial complexes (Figure 14). However, particles of 2.2-2.7 Å radius can make partial complexes in a higher percentage. Particles of radius below 1.4 Å and above 2.8 Å do not form stable full or partial complexes, and rather they show a higher percentage of uncomplexed states in smaller sizes and weakly associated states in larger sizes.

Normalized distributions of the distance (d) between the defined plane of γ-CD and probe particle are shown in Figure 15. With the enlargement of the size, the particles move towards the larger ring uniformly.

The summary of average distance distribution of particles with all three CDs is shown in Figure 16. According to the standard deviation from the average distance, smaller particles have more freedom of movement compared to the larger particles. The same results are shown against the effective radius (eR) in the Figure 17. The formation of CD inclusion complexes directly depends on the dimensions of the CD cavity and the guest molecule. If the guest molecule is too large or bulky, it will not fit completely into the CD cavity and likewise, very small guest molecules will not form stable complexes with CDs as they will slip out of the cavity (Loftsson, 2002).

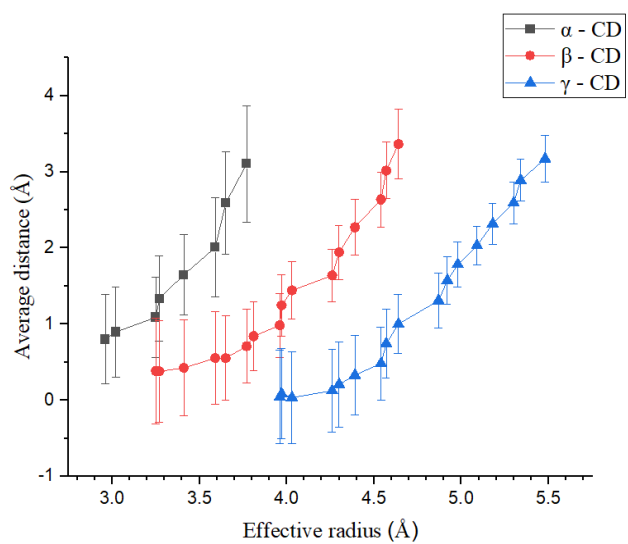

**Figure 17:** Average distance change with the effective radius of the particle.

## CONCLUSION

In this study, the complexation behaviour of cyclodextrin (CD) and guest molecules with 1:1 stoichiometry were extensively studied. All *α, β*, and *γ*-CDs have a high affinity to make stable complexes with a broader size range of probe particles. The complexes formed are in dynamic nature making full and partial complexes since the process is drive by soft, non-covalent interactions. The mean distance from the defined average frame increased uniformly with the increment of the guest particle size and the narrow distribution of distances suggests strong complex formation. Some larger sizes of probe particles showed closely contacted associated behaviour with the wider ring of the CD cavity, even though they could not make full or partial complexes due to the size. The guest particle sizes that CDs can make complexes with are above the approximated cavity sizes in prior work. Since CDs can expand to engulf the particle favourable sizes, this would be possible.

## ACKNOWLEDGEMENT

Financial assistance from the University of Peradeniya Research Grant (URG/2016/55) is acknowledged.

## DECLARATION OF CONFLICT OF INTEREST

The authors declare no competing interest.